\documentclass[preprint]{aastex631}

\usepackage{amsmath}
\usepackage{esint}
\usepackage{xcolor}
\usepackage[colorinlistoftodos]{todonotes}

\let\a=\alpha \let\b=\beta \let\g=\gamma \let\d=\delta \let\e=\epsilon
\let\h=\eta  \let\k=\kappa \let\l=\lambda 
\let\s=\sigma \let\t=\tau \let\u=\upsilon
\let\r=\rho
 
\let\w=\omega	   \let\G=\Gamma \let\D=\Delta  \let\L=\Lambda
\let\S=\Sigma \let\W=\Omega
\let \x = \chi \let \z = \xi

\def\be{\begin{eqnarray}}
\def\ee{\end{eqnarray}}
\def\ba{\begin{array}}
\def\ea{\end{array}}

\def \Msun {M_{\odot}}

\def \max {\mathrm{max}}
\def \min {\mathrm{min}}
\def \sec {\mathrm{sec}}
\newcommand{\hr}{\mathrm{hr}}

\newcommand{\muas}{\mathrm{\mu as}}

\newcommand{\kpc}{\mathrm{kpc}}

\def \M  {\mathscr{M}}
\def \K {\mathrm{K}}

\newcommand{\ghz}{\mathrm{GHz}}
\newcommand{\gcc}{\mathrm{g\ cm^{-3}}}
\newcommand{\jy}{\mathrm{Jy}}
\newcommand{\mjy}{\mathrm{mJy}}

\def \earth {\oplus}
\def \S {Section }

\newcommand{\sobs}{\sigma_\mathrm{obs}}
\newcommand{\afine}{c_\mathrm{fine}}
\newcommand{\acomb}{c_\mathrm{comb}}

\newcommand{\rg}{r_\mathrm{g}}
\newcommand{\tg}{t_\mathrm{g}}
\newcommand{\pmag}{p_\mathrm{mag}}
\newcommand{\betaw}{\beta_\mathrm{w}}
\newcommand{\tti}{T_\mathrm{i}}
\newcommand{\tte}{T_\mathrm{e}}
\newcommand{\rlow}{R_\mathrm{low}}
\newcommand{\rhigh}{R_\mathrm{high}}

\shorttitle{Stellar wind fed accretion reproduces Sgr~A* variability}
\shortauthors{Murchikova, White, \& Ressler}

\graphicspath{{./}{figures/}}

\begin{document}

\title{Remarkable correspondence of Sagittarius A* submillimeter variability\\with a stellar-wind-fed accretion flow model}

\correspondingauthor{Authors' contributions are equivalent}
\email{lena@ias.edu, cjwhite@princeton.edu, smressle@ucsb.edu}

\author[0000-0001-8986-5403]{Lena Murchikova}
\affiliation{Institute for Advanced Study, 1 Einstein Drive, Princeton, NJ 08540, USA}

\author{Christopher J. White}
\affiliation{Department of Astrophysical Sciences, Princeton University, Peyton Hall, Princeton, NJ 08544, USA}

\author[0000-0003-0220-5723]{Sean M. Ressler}
\affiliation{Kavli Institute for Theoretical Physics, University of California Santa Barbara, Kohn Hall, Santa Barbara, CA 93107, USA}

\begin{abstract}

We compare the 230 GHz near-horizon emission from Sagittarius A* to simulations representing three classes of accretion flows. Using the structure function to capture the variability statistics of the light curve, we find a noticeable discrepancy between the observations and models based on torus-fed accretion disks, whether those disks bring in a small or large amount of net magnetic flux. On the other hand, the simulations that are fed more realistically by stellar winds match the observed structure function very well. We describe the differences between models, arguing that feeding by stellar winds may be a critical component in constructing theoretical models for accretion in the Galactic Center.

\end{abstract}


\section{Introduction}

Sagittarius~A* (Sgr~A*) is the closest supermassive black hole to us, and is our best hope to study near-horizon phenomena. The peak of its electromagnetic emission is in submillimeter (submm) range \citep[around $230\ \ghz$,][]{Genzel2010} and this emission is currently one of the key probes of the near-horizon physics (via, e.g., the Event Horizon Telescope, \citealt{EHTI}) and hence understanding its properties is critically important for the whole field to move forward. The submm emission is particularly useful for this purpose because other observational probes of the near-horizon physics (e.g., the NIR \citealt{Gravity2019,Abuter2020} emission and X-rays \citealt{Neilsen2015}) are likely connected to high-energy, nonthermal particle acceleration, a process that involves small-scale plasma physics not captured in global accretion simulations.

Two important and complimentary ways to analyze this submm emission are variability studies and black hole imaging studies. Imaging highlights spatial information while typically averaging data in time. Variability analyses highlight temporal evolution while often discarding spatial information. Therefore, in order to obtain a complete picture of black hole/accretion physics, both sets of analysis are required.

The behavior of Sgr~A*'s submm flux is not yet fully explained by models. In this work we analyze the submm variability of Sgr~A* on scales between about 10 seconds to about 200 minutes. We compare observational results \citep{Murchikova2021b,Witzel2020,Dexter2014} with simulations \citep{White2020, Ressler2020}, and identify an accretion flow model that describes Sgr~A* variability nearly perfectly. This model traces accretion from stellar winds at parsec scales down to the event horizon.

Our numerical models include 
\begin{itemize}
\item[(i)] SANE torus. A Standard and Normal Evolution (SANE) accretion flow fed by torus initial conditions from \citet{White2020}, with a magnetic field with no large-scale net vertical flux.
\item[(ii)] MAD torus. A Magnetically Arrested Disk (MAD) accretion flow fed by torus initial conditions, with a large net vertical flux in the inner region.
\item[(iii)] MAD winds. Two accretion flows fed by stellar winds from \citet{Ressler2020}, both of which naturally become MAD.
\end{itemize}

As a measure of Sgr~A* variability we use the intrinsic structure function ($\mathcal{SF}$) defined in \cite{Murchikova2021b} as an extension of the standard structure function definition to the case where observational uncertainties are present:
\begin{equation}
    \mathcal{SF}(\tau)=\sqrt{\frac{1}{N_\tau}\sum\limits_{\mathrm{pairs}}(F(t+\tau)-F(t))^2 - 2 \sobs^2}.
    \label{eq:sf}
\end{equation}
Here $\tau$ is the time lag, $F(t)$ is the flux at time $t$, $N_\tau$ is the number of pairs of points separated by the time interval $\t$ in the data, the summation runs over all such pairs, and $\sigma_\mathrm{obs}$ is the root-mean-square observational uncertainty. In the case of the simulated variability the observational uncertainties are absent ($\sigma_\mathrm{obs}=0$). For brevity, throughout the text, we simply refer to $\mathcal{SF}$ as the structure function.

We describe the observational data set and construction of the structure function in Section~\ref{sec:observations}. In Section~\ref{sec:simulations}, we cover the simulations in more detail. Section~\ref{sec:sim_lightcurves} describes the process by which we extract submm variability from the simulations. Our analysis follows in Section~\ref{sec:discussion}, and we conclude in Section~\ref{sec:conclusion}. In Appendix \ref{app:model}, we demonstrate properties of structure functions with simple models.

\section{Observational Data}
\label{sec:observations}

We use observational data sets from \citet{Murchikova2021b}, \citet{Witzel2020}, and \citet{Dexter2014}. The data were obtained at the frequency of about $230\ \ghz$ with the Atacama Large Millimeter/submillimeter Array (ALMA), the Submillimeter Array (SMA), and Combined Array for Research in Millimeter-wave Astronomy (CARMA) between 2009 and 2019. Detailed descriptions of the observations and the data reduction can be found in the corresponding papers. We reanalyze the ALMA data from project 2018.1.01124.S (PI Murchikova) described in \citet{Murchikova2021b}. We update the light curve extraction procedure, which allows us to obtain the timestamps with decimal second accuracy. The rest of data analysis and the data reduction are unchanged.

To construct the structure function at timescales $\tau\leq 200$ seconds we use only the lowest-noise part of the \citeauthor{Murchikova2021b} data set. It contains the five (out of seven) best ALMA observations with average uncertainty of $\sobs^\mathrm{low}=4\ \mjy$. At $\tau \sim 200\ \sec$ the value of the structure function $\mathcal{SF}$ becomes greater than ${\sim}3 \sobs^\mathrm{high}$, where $\sobs^{high}=13\ \mjy$ is the observational uncertainty of the higher noise part of the \citeauthor{Murchikova2021b} data set. For $200\ \sec < \tau < 2000\ \sec$ we use their entire data set. The upper threshold is chosen to be where we start to become limited by statistics. It is equal to about one-half the length of ALMA execution, ${\sim}70/2=35\ \min$, as we only have 14 independent samples of this length in this data. 

On time scale between 1400 seconds (about 25 minutes) and 11,000 seconds (190 minutes) we use a combination of \citet{Witzel2020} and \citet{Dexter2014} data sets. The lower threshold is determined by the time scale at which the the value of the structure function calculated from the data is greater than the few sigmas of observational uncertainties per data set used. The upper threshold is chosen such that data points from at least two telescopes (SMA and CARMA at such timescales) are present, and that there are about 10 independent data stretches of $\tau$ length in each of these telescopes.

To construct the average structure function on timescales $25\ \min \leq \tau \leq 190\ \min$ we calculate the $\mathcal{SF}$ using Equation~\ref{eq:sf} for data from each telescope individually and then average them. The total lengths (with gaps removed) of ALMA and SMA data sets in \citet{Witzel2020} and SMA and CARMA data sets in \citet{Dexter2014} are about ${\sim}2000\ \min$, therefore we average the structure functions with identical weights. We choose this averaging to preserve the physical meaning of the structure function, which is the offset of the flux value $F(t+\tau)$ from $F(t).$ This approach also allows us to avoid high-variability and high-noise measurements (particularly CARMA) dominating the rest of the sample.

The long-time-scale part of the $\mathcal{SF}$ is multiplied by a factor 0.87 to make it aligned with the short-time-scale part of the structure function. If we remove the CARMA data set (the noisiest and the most variable data set), the long-time-scale part of $\mathcal{SF}$ would have to my multiplied by a factor of 1.05 and the shape of the curve would not change. Both sets of points are plotted in Figure \ref{fig:sf_comparison} and are barely distinguishable. 

There is a general trend we observe in Sgr~A* structure functions calculated with different data sets:\ the shapes of the $\mathcal{SF}$ are the same, but they tend to be offset from each other by some constant multiplicative factor varying from observation to observation.

Our investigation provides additional support for wind-fed modelling being appropriate for the Galactic Center, and it motivates the further study of this class of theoretical models.

\section{Simulations}
\label{sec:simulations}

The simulations we use are all evolved with \texttt{Athena++} \citep{Stone2020}, using its ideal general-relativistic magnetohydrodynamics (GRMHD) capabilities \citep{White2016}. Here we provide details in terms of the gravitational length $\rg = G M / c^2$ and time $\tg = G M / c^3$. In case of the Sgr~A* black hole, the gravitational length is about $0.04$ AU and the gravitational time is about $20$ seconds.

\subsection{SANE Torus}

For the SANE case, we use the aligned simulation first described in \citet{White2020}, similar to the standard SANE models studied in the literature and compared in \citet{Porth2019}. The black hole has dimensionless spin $a = 0.9$, and the initial conditions are a prograde hydrodynamic equilibrium torus from \citet{Fishbone1976} with inner edge at $r_\mathrm{in} = 15\ \rg$ and pressure maximum at $25\ \rg$. The fluid is taken to have an adiabatic index of $\Gamma = 4/3$. An initial poloidal magnetic field is added to the torus, normalized such that the density-weighted average of plasma $\beta^{-1}$ (the ratio of magnetic pressure $\pmag$ to thermal pressure) is $0.01$.

The simulation coordinates are spherical Kerr--Schild, where three levels of static mesh refinement beyond the root grid achieves an effective resolution of $448 \times 256 \times 352$ cells in radius, polar angle, and azimuthal angle within $50^\circ$ of the midplane. Radial spacing is logarithmic ($239$ cells per decade), running from inside the horizon to $r = 100\ \rg$.

The simulation is evolved to a time of $t = 11{,}000\ \tg$, reaching inflow equilibrium beyond $r = 20\ \rg$. Only data past $5000\ \tg$ is used for this analysis. The dimensionless magnetic flux $\varphi$ saturates between $7$ and $17$ in Gaussian units, well below the MAD regime of approximately $47$ \citep{Tchekhovskoy2011}. After fixing the density scale in order to match the overall observed submm flux (see Section~\ref{sec:sim_lightcurves}), the average density in the innermost $10\ \rg$ is $2 \times 10^{-17}\ \gcc$.

\subsection{MAD Torus}

For the MAD case, we employ a standard $a=0.9375$ MAD torus simulation like those commonly studied in the literature \citep[e.g.,][]{McKinney2012}. The initial \citealt{Fishbone1976} torus has an inner radius of $r_\mathrm{in} = 20\ \rg$ and a pressure maximum of $41\ \rg$. The initial magnetic field is set via the vector potential $A_\varphi \propto \max(q,0),$ with
\begin{equation}
    q = \frac{\rho}{\rho_\mathrm{max}} \left(\frac{r}{r_\mathrm{in}}\right)^3 \sin^3(\theta)\exp\left(-\frac{r}{400\ \rg}\right) - 0.2,
\end{equation}
where $\rho$ is the fluid-frame mass density, $\rho_\mathrm{max} = 1$, and the proportionality constant is set so that the maximum thermal pressure in the torus divided by the maximum of $\pmag$ in the torus is 100.

Small perturbations are added to the initial torus pressure at the 2\% level. The adiabatic index for this simulation is $\Gamma = 13/9$. The simulation is evolved for $10{,}000\ \tg$, with the last $5000\ \tg$ used for analysis. Here, the density within $10\ \rg$ of the center is $4\text{--}5 \times 10^{-19}\ \gcc$.

\subsection{MAD Stellar Winds}

Finally, we consider the two stellar wind simulations from \citet{Ressler2020}. Both of these model accretion onto a nonspinning black hole of matter sourced by realistic winds from the approximately $30$ Wolf--Rayet stars nearest to the Galactic Center. While the orbits and mass-loss rates of these stars are well known, the magnetic structure of their winds is more uncertain. One simulation assumes the plasma $\betaw$ of the winds is $10^2$;\ the other assumes less magnetization with $\betaw = 10^6$. Here, $\betaw$ is defined as the ratio of ram pressure $\rho v^2$ to $\pmag$ in the wind.

These simulations use Cartesian Kerr--Schild coordinates. The root grid extends to $1600\ \rg$ in each direction, with $128^3$ cells. Nine nested levels of static mesh refinement are added, with $128^3$ cells covering the inner $(6.25\ \rg)^3$ at the highest level.

Both wind-fed simulations are run for a time of $20{,}000\ \tg$, with a steady state region (defined by an accretion rate independent of radius) extending to approximately $r = 100\ \rg$. The latter $10{,}000\ \tg$ is used in this analysis. Both cases result in MAD flows, with $40 \lesssim \varphi \lesssim 60$. The density inside $r = 10\ \rg$, set by the known stellar winds themselves, averages $4.5 \times 10^{-19}\ \gcc$ ($\betaw = 10^2$) and $4.2 \times 10^{-19}\ \gcc$ ($\betaw = 10^6$).

\begin{figure*}
\includegraphics[width=0.97\textwidth]{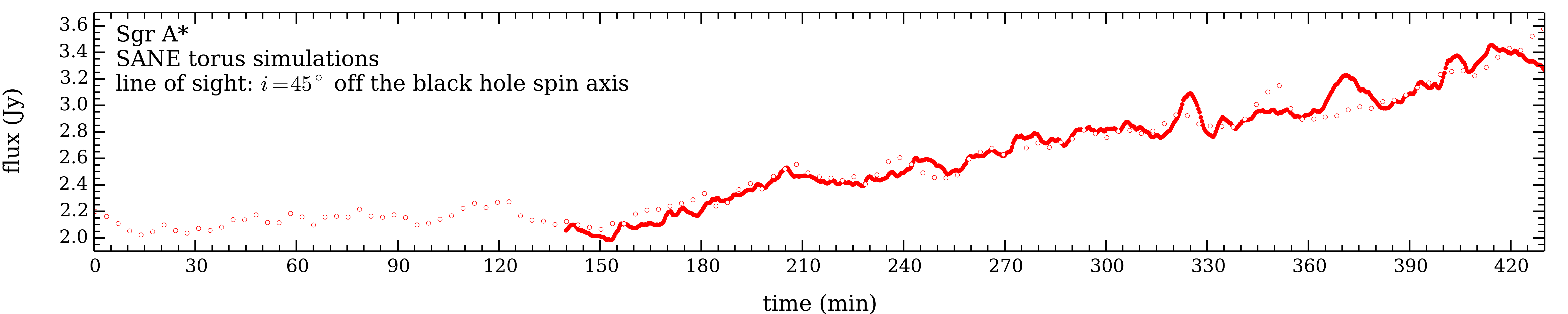}\\
\includegraphics[width=0.97\textwidth]{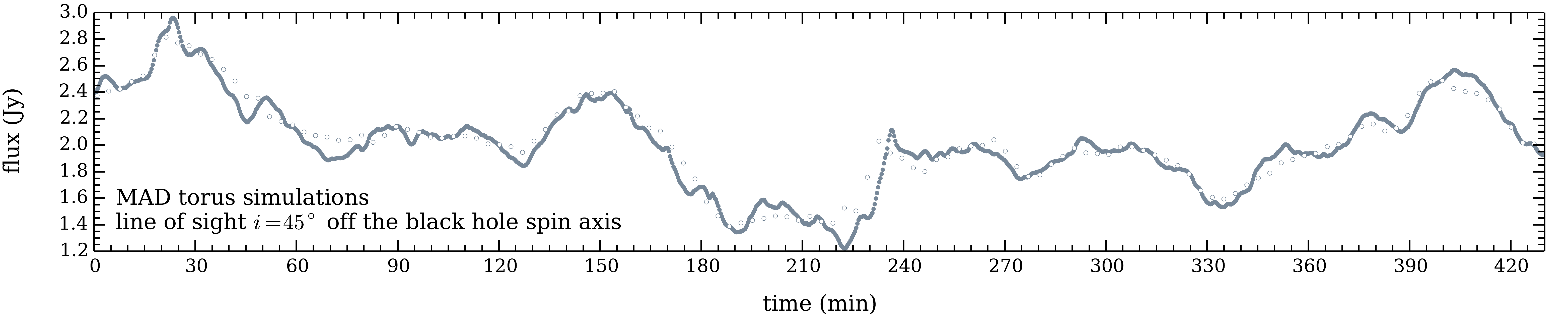}\\
\includegraphics[width=0.97\textwidth]{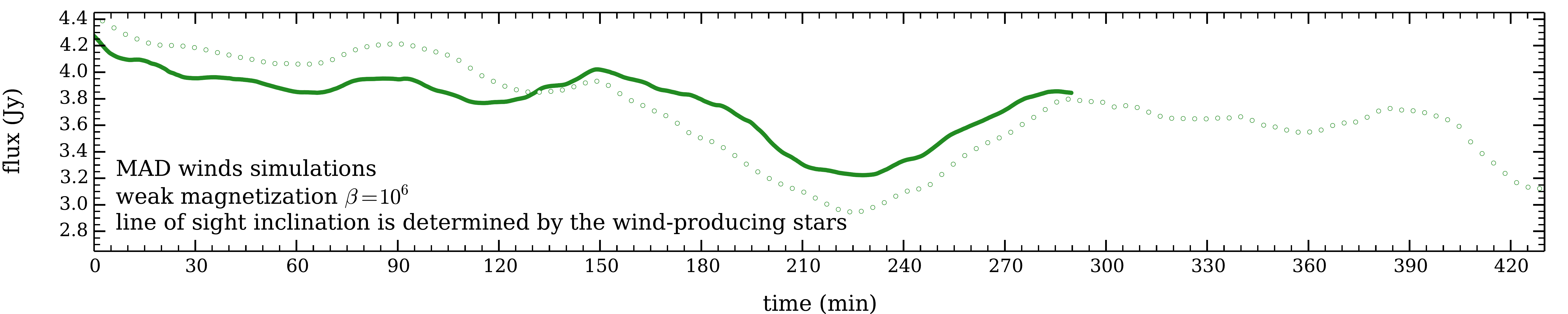}\\
\includegraphics[width=0.97\textwidth]{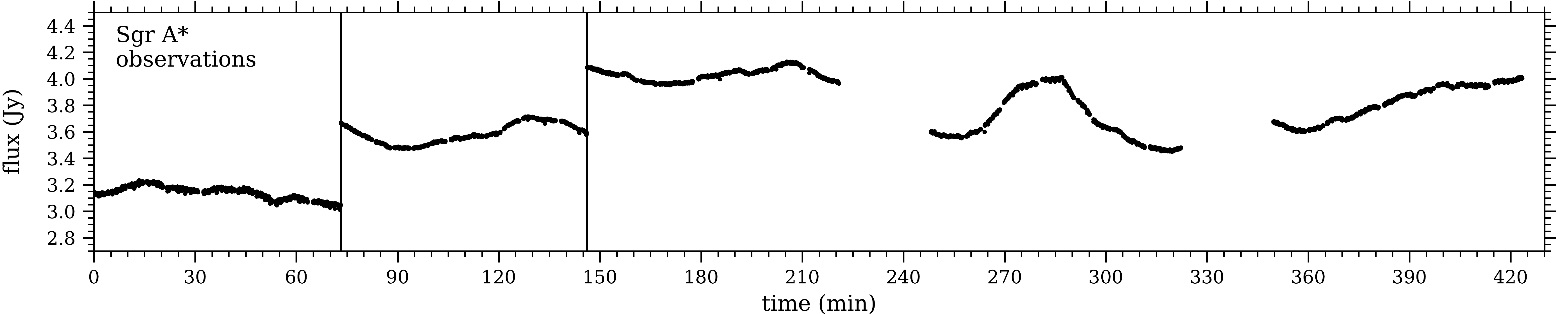}
\caption{Sgr~A*'s simulated and observed light curves. Sample simulated data for SANE torus, MAD torus and MAD winds is plotted in red, grey, and green, respectively. Solid points represent finely sampled high-cadence light curves, and empty points represent coarsely sampled data obtained with the fast-light approximation (Section~\ref{sec:sim_lightcurves}). Sample observational data from \citealt{Murchikova2021b} is plotted in black.}
\label{fig:lightcurves}
\end{figure*}

\begin{figure}
\begin{centering}
\includegraphics[width=0.32\columnwidth]{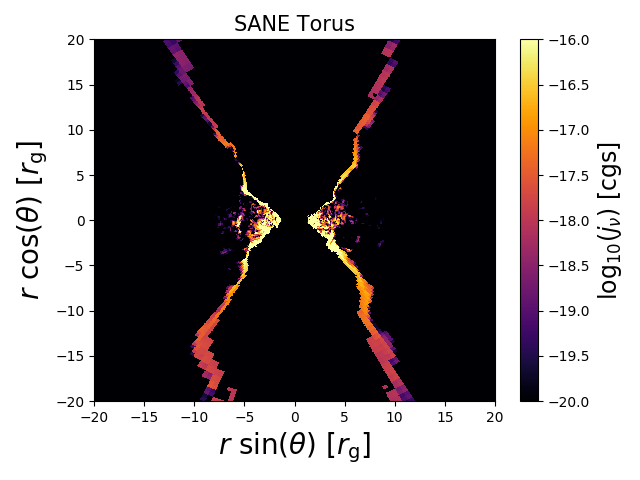}
\includegraphics[width=0.32\columnwidth]{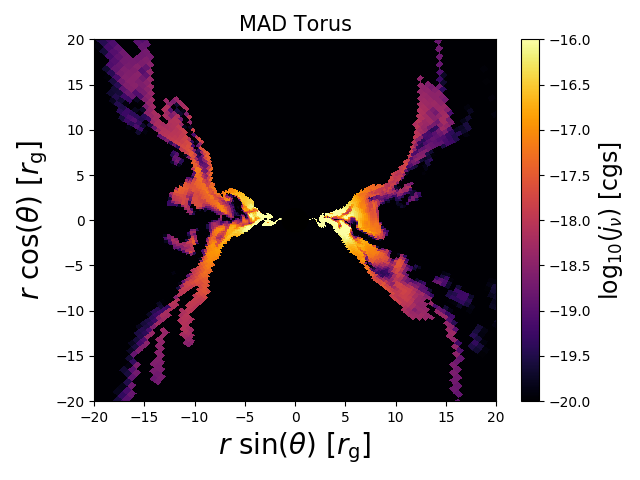}
\includegraphics[width=0.32\columnwidth]{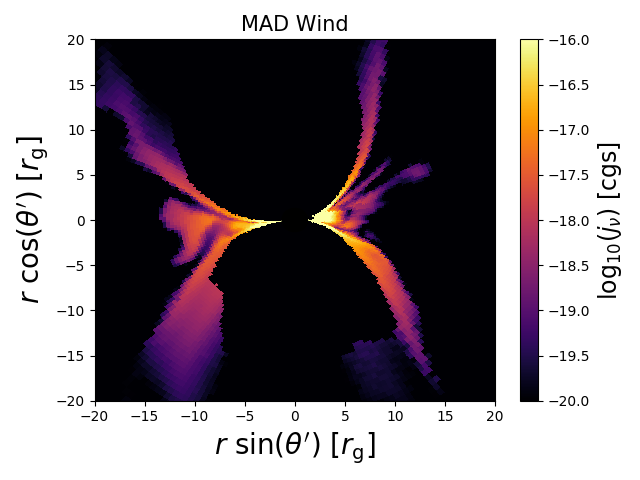}
\caption{Approximate $230\ \ghz$ synchrotron emissivity, $j_\nu$, calculated using Equation (57) of \citet{Leung2011} for the three simulations used in this work. Regions with plasma $\sigma \equiv 2 \pmag / (\rho c^2) > 1$ are assigned zero emissivity. We reorient the angle of viewing such that the flows are similarly aligned; in all cases the average direction of the angular momentum vector is pointing upward.
Left: SANE torus simulation.
Middle: MAD torus simulation.
Right: $\betaw=10^2$ MAD winds simulation. 
Both MAD simulations have more extended emitting regions than the SANE torus simulation. Furthermore, the MAD torus displays more small-scale turbulence in the emitting regions than the MAD winds. The light curves in Figure~\ref{fig:sf_comparison} reflect these differences, with both MAD simulations displaying longer-term variations than the SANE simulation. Additionally, the MAD winds light curve is much smoother than the other light curves because of the comparative lack of small-scale turbulence.}
\label{fig:jnu}
\end{centering}
\end{figure}

\section{Extracting submm variability from simulated data}
\label{sec:sim_lightcurves}

For all simulations, we use the general-relativistic ray-tracing code \texttt{grtrans} \citep{Dexter2016,Dexter2009} to produce $230\ \ghz$ images from simulation snapshots. These images account for polarized synchrotron emission, absorption, and rotation/conversion. Each image is made with $256^2$ pixels covering a field of view of $24\ \rg$ ($120\ \muas$) on each side, where we take the black hole to have mass $M = 4.152 \times 10^6\ \Msun$ and to be at a distance of $8.178\ \kpc$ \citep{Gravity2019}. Only a ${\sim}5\%$ change would be induced by instead using the values obtained by \citet{Do2019}:\ a mass of $3.964 \times 10^6\ \Msun$ and a distance of $7.946 \ \kpc$. Integrating the intensity over an image yields a simulated light-curve data point.

For the stellar wind models, the ray-tracing camera is positioned to correspond with the line of sight from Earth. We use the last $1000$ simulation snapshots, spaced by $10\ \tg$ ($200\ \sec$) and thus covering a time of $56.8\ \hr$. Though ideal GRMHD simulations are scale free (the fluid density, fluid pressure, and magnetic field strength can all be scaled up or down consistent with one another), these simulations fix this degree of freedom by knowing the absolute mass-loss rates from the Wolf--Rayet stars. The only freedom we have in generating images is our choice of how electron temperature (which is not important for the hydrodynamics but critical for synchrotron radiation) is determined from the total fluid temperature and other variables. We use the standard prescription from \citet{Moscibrodzka2016}, which sets the ion-to-electron temperature ratio based on plasma $\beta$:
\begin{equation}
  \frac{\tti}{\tte} = \frac{\rlow + \beta^2 \rhigh}{1 + \beta^2}.
\end{equation}
We fix $\rlow = 1$ and tune $\rhigh$ until the average flux is $2.4\ \jy$, finding reasonable values of $23$ ($\betaw = 10^2$) and $9.2$ ($\betaw = 10^6$).

We use a similar procedure for the torus simulations, though in these cases we have more degrees of freedom. Without knowing the orientation (if any) of angular momentum around Sgr~A*, the viewing angle is a free parameter, and so we make images from both $5^\circ$ and $45^\circ$ off the gas angular momentum (and black hole spin) axis. Additionally, we must choose a physical scale for the model. For this we fix $\rhigh = 16$ (intermediate between the two stellar wind values) and adjust the physical scale independently for each viewing angle until the average flux is again $2.4\ \jy$. For these simulations, we use the last $601$ (SANE) or $501$ (MAD) snapshots, again separated by $10\ \tg$ in time, covering $34.1\ \hr$ and $28.4\ \hr$, respectively.

The ray tracing we have discussed so far employs the fast-light approximation, where a single simulation snapshot is used to make an image, assuming no quantities change while light propagates through the system. Though this is adequate for timescales roughly $10\ \tg$ or longer, light curve properties (especially variability) on shorter timescales might be influenced by the fact that the emitting matter is moving at relativistic speeds and so cannot be stationary over a light-crossing time. We therefore also generate images from high-cadence simulation dumps ($\Delta t = 1\ \tg = 20\ \sec$), employing the slow-light capability of \texttt{grtrans}, which uses multiple dumps simultaneously to account for the evolution of the simulation while photons are propagating. We keep the tuned physical scaling and $\rhigh$ values already found. In this way we produce high-cadence light curves with $852$ (SANE torus), $4852$ (MAD torus), and $851$ (MAD wind) samples, covering $4.8$, $27.6$, and $4.8\ \hr$, respectively. The resulting light curves are presented in Figure~\ref{fig:lightcurves}.

To obtain the variability structure functions from model light curves, we combine the structure functions calculated using Equation~\ref{eq:sf} for coarsely and finely sampled data independently. The coarsely sampled light curves are generally longer and consequently better sample the absolute amplitude. Moreover, slow light calculations require the camera not be too far from the source (lest the calculation become prohibitively expensive), possibly missing some (essentially constant) emission originating from larger radii. Thus, we multiply the fine-sampling structure function by a coefficient $\afine$ to align it with the coarse-sampling structure function.
The alignment coefficients are are as follows: for the SANE torus with $i = 5^{\circ}$ or $45^\circ$, $\afine = 0.9$ or $0.78$, respectively; for the MAD torus with the same inclinations, $\afine = 0.86$ or $0.8$; for the MAD winds, $\afine$ is either $1.2$ ($\betaw = 10^2$) or $0.94$ ($\betaw = 10^6$).

To determine whether the orientation of the flow in the case of MAD winds model significantly affects the appearance of the light curves and the structure functions, we conduct an inclination test. For the $\betaw = 10^6$ MAD wind simulation we mock-observe the black hole from both the physical viewing angle (determined by the stellar wind feeding) and at an artificial $45^\circ$ inclination. We find that this change has a negligible effect on the structure function.

\section{Discussion}
\label{sec:discussion}

\begin{figure*}
\vspace{-0.0cm}
\centering
\begin{tabular}{ccc}
\includegraphics[width=0.31\textwidth]{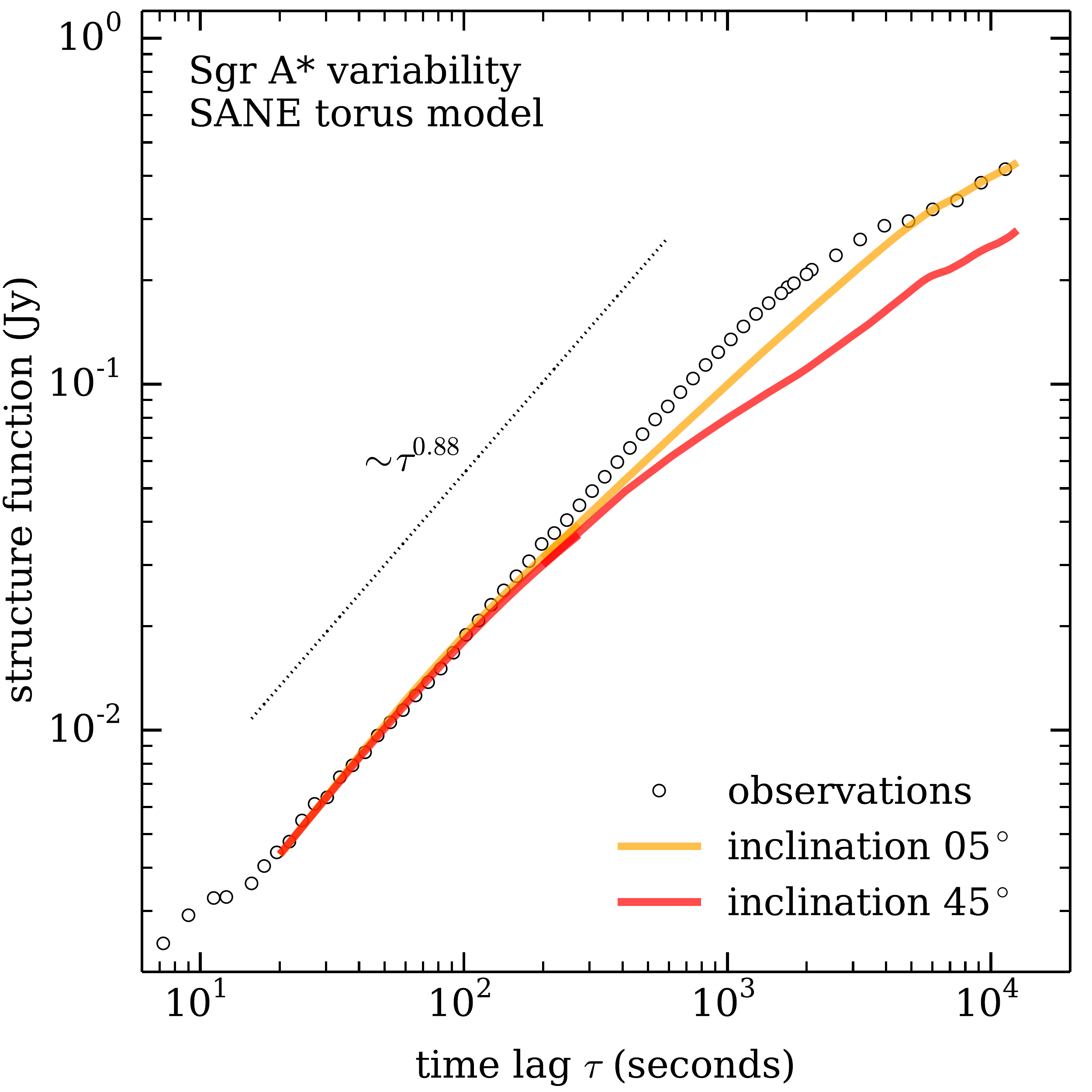} &
\includegraphics[width=0.31\textwidth]{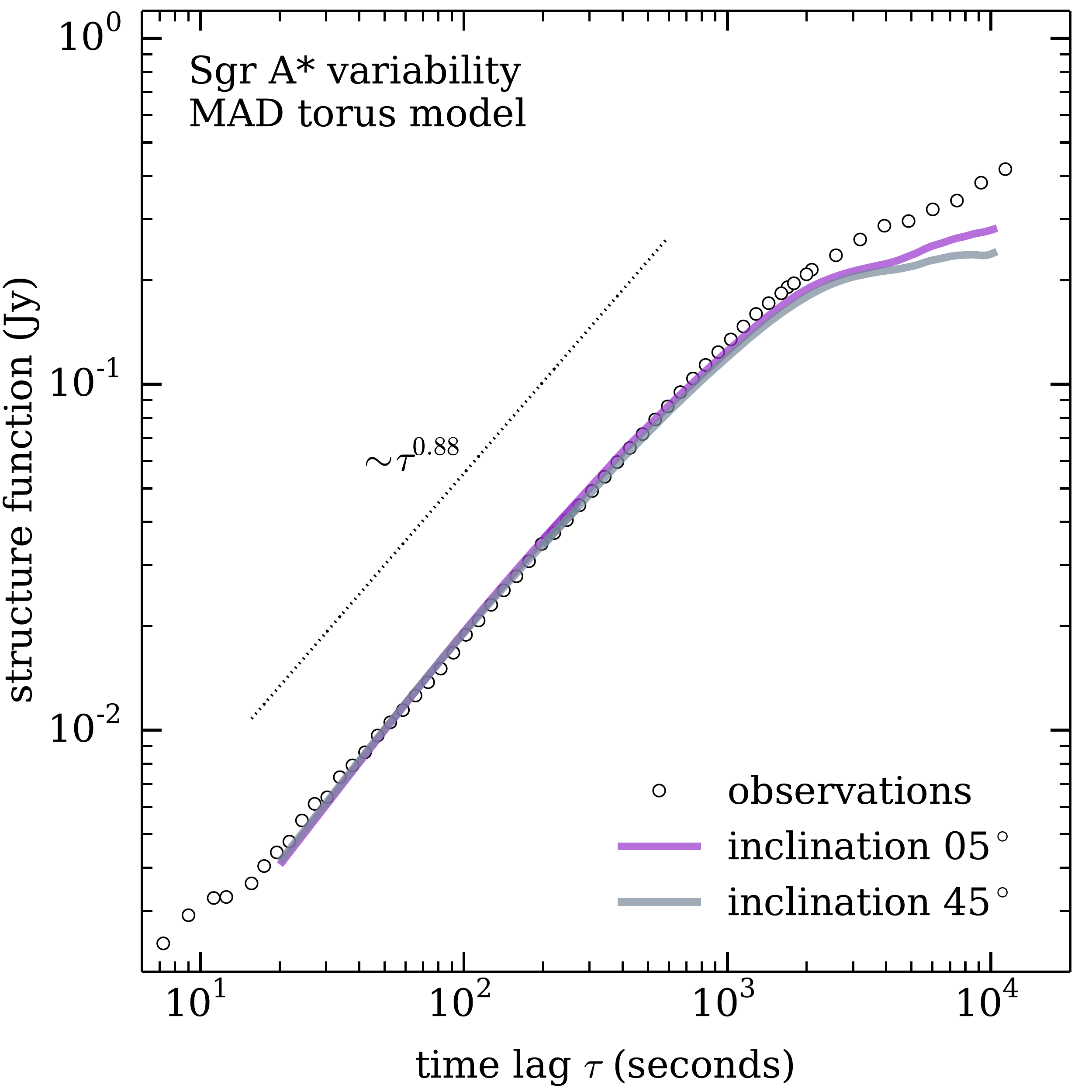} &
\includegraphics[width=0.31\textwidth]{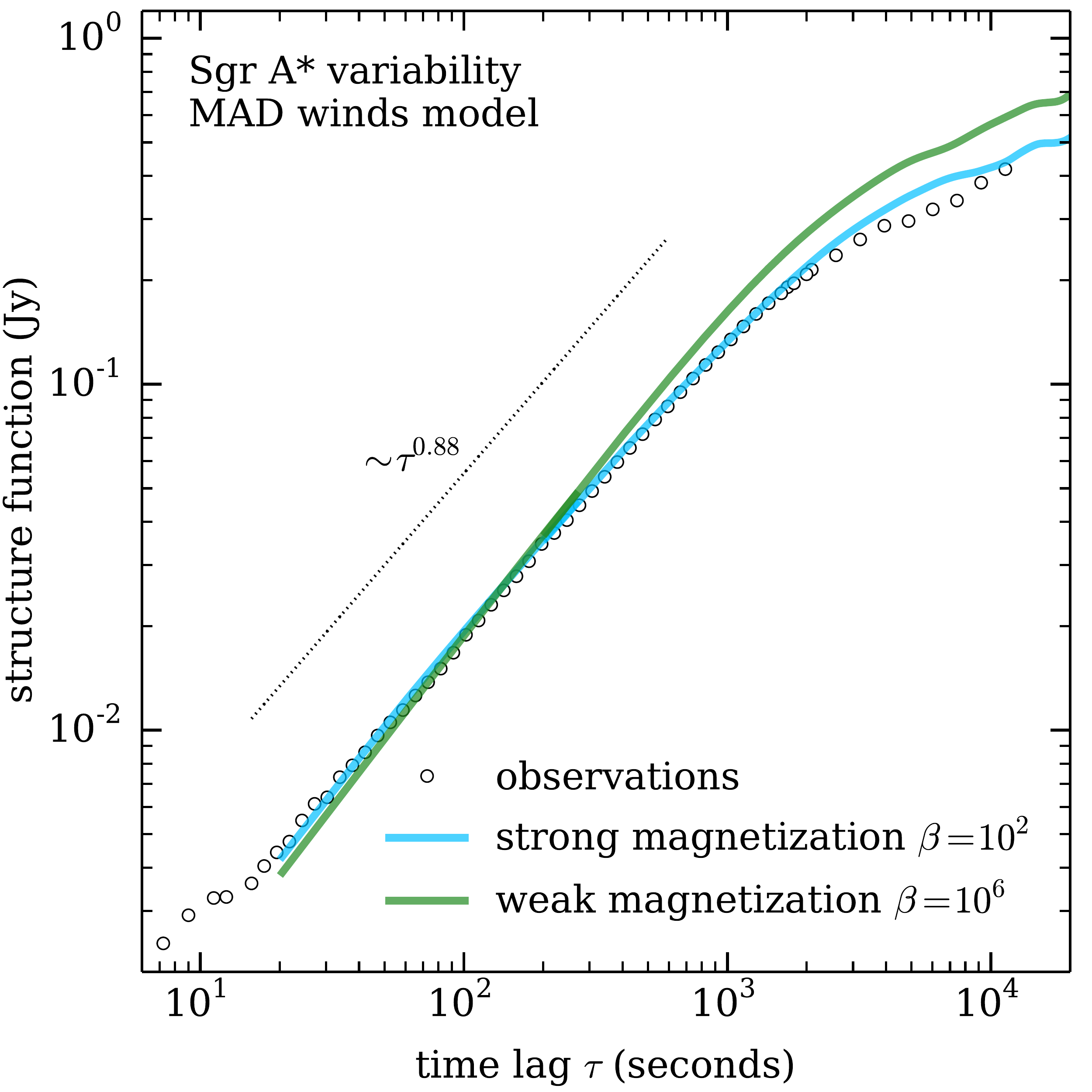}
\end{tabular}
\caption{Variability structure function of Sgr~A* at $230~\ghz.$ Comparison of simulations and observations. The structure function obtained from observed light curves is depicted with empty black circles.
Left:\ simulations of SANE Sgr~A* accretion flow with torus initial conditions as seen from $5^\circ$ (yellow) and from $45^\circ$ (red) off the spin axis of the black hole.
Middle:\ simulations of MAD Sgr~A* accretion with torus initial conditions as seen from $5^\circ$ (purple) and from $45^\circ$ (grey) off the spin axis of the black hole.
Right:\ MAD stellar-wind-fed Sgr~A* accretion simulations with strong magnetization $\betaw = 10^2$ (blue) and weak magnetization $\betaw = 10^6$ (green) of the stellar winds.
The simulated structure functions are normalized by coefficients $\acomb$ to align with the observed values. The coefficients are as follows: for the SANE torus, $\acomb = 0.72$ ($i = 5^\circ$) or $0.6$ ($i = 45^\circ$); for the MAD torus, $0.45$ or $0.4$ for the same inclinations; for the MAD winds, $\acomb = 0.7$ ($\betaw = 10^2$) or $1.15$ ($\betaw = 10^6$).
}
\label{fig:sf_comparison}
\end{figure*}

By-eye comparison of the simulated and observed light curves in Figure~\ref{fig:lightcurves} leads us to conclude that the stellar-wind-fed accretion model (MAD winds) produces light curves much closer in shape to the observed one, compared to either the SANE torus or the MAD torus. The typical snapshot of a MAD winds flow looks smoother and less stochastic than other analysed models (Figure~\ref{fig:jnu}).

To quantify the variability properties we use the structure function $\mathcal{SF}(\tau)$ defined by Equation-\ref{eq:sf}. The physical meaning of this quantity is the root-mean-square of the deviation of the black hole flux between moments in time separated by $\tau.$ The comparison of the models with the observational data is presented in Figure~\ref{fig:sf_comparison}.

At small timescales all our structure functions follow the power-law slope of ${\sim}\tau^{0.88}.$ This is close to the universal slope of ${\sim}\tau^{1.0}$ derived in Appendix~\ref{app:model}. The universal slope, $\mathcal{SF} \sim\tau^{1.0}$, appears when we study structure function behavior on time scales much smaller than the smallest period of oscillations of the electromagnetic emission in the source. The closeness of the observed index 0.88 to 1.0 indicates that this condition is nearly met. 

The fact that the power-law slope is 0.88 and not 1.0 indicates that weak modes with short periods are present in the data. This is particularly interesting in the case of timescales about 10 seconds, which is smaller than the time it takes light to circle halfway around the black hole as fast as possible (via the prograde photon orbit), $6\text{--}16\ \tg = 130\text{--}330$ seconds, depending on spin. Such short timescales might arise naturally from brief bursts of emission from small-scale turbulence, though this is less applicable to the wind-fed models (Figure~\ref{fig:jnu}). They may be indicative of relativistic velocities moving emitting regions across parts of the domain that are highly magnified by gravitational lensing.

Figure~\ref{fig:sf_comparison} shows that the simulated model structure functions trace the observed one with different degrees of success. The SANE torus model deviates from the observed curve for $\tau \gtrsim$ 150 seconds (2.5 minutes). The MAD torus model successfully traces the observations until $\tau \approx$ 1000 seconds (about 15 minutes) while the slope remains ${\sim}\tau^{0.88}.$ Then the observed slope starts gently flattening at larger $\tau$, however the simulated structure function flattens much faster, and thus deviates substantially.

The MAD winds are the most successful models. The simulated curves trace the observed structure function almost perfectly. They follow the observed slope ${\sim}\tau^{0.88}$ until about 1000 seconds, then flatten as gently as in the observations. The model with strong magnetization of the stellar winds at the source ($\betaw=10^2$) does particularly well, tracing the observed structure function between 20 seconds (the shortest cadence in our simulations) all the way to about 200 minutes.

The fact that the structure function produced by the MAD winds model follows the observed Sgr~A* structure function between 20 seconds and 200 minutes demonstrates that the properties of the light curves in the model and the physical system are similar. In particular, the harmonics influencing the source variability on these timescales are nearly identical. Deviation of the structure functions produced by SANE torus and MAD torus models from the observed curve on shorter timescales strongly indicates the presence of modes with shorter periods, which are absent in the observations of Sgr~A* (see Appendix~\ref{app:model}). These higher frequency variations are connected to smaller scale turbulence, as evident from the 2D contour maps of the 230 GHz emissivity shown in Figure~\ref{fig:jnu}. 

SANE torus, MAD torus, and MAD winds models are evolved using \texttt{Athena++} with the same underlying plasma physics. The crucial difference is the black hole feeding mechanism. Torus models feed the back hole through initializing a torus on the scale of tens to hundreds of gravitational radii, with the understanding that after sufficient evolution and inside a sufficiently small radius, the resulting accretion disk will be in a quasi-steady state that is independent of the details of those initial conditions. However, the physical system cannot so easily forget the initial property of all material moving on circular orbits with the same orientation of angular momentum. The MAD wind simulations, on the other hand, include fluid parcels with eccentric orbits with many different orientations, reaching a qualitatively different quasi-steady state. Crucially, it is the latter case that reflects, by construction, the actual feeding of gas in the Galactic Center from ${\sim}30$ Wolf--Rayet stars.

From our analysis it is evident that the black hole accretion flow is sensitive to the precise feeding mechanism. When we take care to include realistic feeding in simulations, the predicted $230\ \ghz$ emission reproduces the observed variability of Sgr~A* (as in the third and the fourth panels in Figure \ref{fig:lightcurves}). Phenomenological torus models, on the other hand, predict $230\ \ghz$ variability that is too noise-like (i.e., high frequency, low amplitude), with short timescale variability not seen in the observations (as in the first and second panels in Figure~\ref{fig:lightcurves}).

Wind-fed GRMHD modelling of black hole accretion at horizon scales is relatively new, and there is a need for further exploration of these sorts of flows with additional simulations, especially given how well they can match the statistical properties seen in Sgr~A*. One parameter we have not fully explored here is that of spin. Currently, the spin of Sgr~A* is essentially unconstrained, and our two wind-fed models assume the black hole is nonspinning. Future work can vary the magnitude and direction of the spin, especially if forthcoming Event Horizon Telescope resolved images narrow the allowed parameter space.

Taking into account other sources of accreting gas, such as the minispiral and Circumnuclear disk in the Galactic Center \citep{Genzel2010}, would also lead to further improvement in accuracy of the numerical models. In general, we expect that the better we model the actual feeding of gas onto the Galactic Center black hole (Solanki et al., in preparation), the more realistic will be the behavior obtained in simulations.

Beyond matching the particulars of Sgr~A*, wind-fed models are worth investigating in order to catalog the ways in which they differ from the torus-initialized simulations that have dominated the literature to date. Stellar winds may be applicable to many low-luminosity supermassive black hole accretion flows. Even without resolved images, we have demonstrated here that light curve statistics can distinguish between these classes of models.

\section{Conclusions}
\label{sec:conclusion}

We compare the observed variability of the Sgr~A* black hole at about $230\ \ghz$ with those generated by three classes of theoretical models for the accretion flow at horizon scales. We rely on the structure function to compare the statistical properties of the light curves. This allows us to compare variability properties without being concerned with particular realizations of stochastic processes.

We consider three types of accretion flow models: SANE torus, MAD torus, and MAD winds. The first two types are torus-initialized models with either a small amount of accreted net magnetic flux or a significant coherent accreted magnetic flux, respectively. The last type consists of stellar-wind-fed models informed by realistic feeding of material onto the black hole by nearby Wolf--Rayet stars. It also results in a MAD accretion flow.

We find that our stellar-wind-fed models match Sgr~A* variability properties much better than either of torus-initialized models. They produce light curves with structure functions nearly identical to the one obtained from observational data. In our analysis we used no parameter tuning to achieve such an excellent match (other than choosing $\rhigh$ so that the 230 GHz flux $\approx$ 2.4 Jy). In general, stellar-wind-fed models have much less room for after-the-fact parameter tuning than torus models; the fact that \emph{any} choice of $\rhigh$ can reproduce the observed amount of Galactic Center horizon-scale synchrotron emission is itself very much non-trivial.

This work lead us to two primary conclusions. First, we have demonstrated that wind-fed models can be distinguished from torus-fed models. This can be done without resolved images of the accretion flow. In fact, it is not a priori guaranteed that time-averaged images of these models are observationally different. 
Therefore, studying time variability is complimentary to image analysis and gives us another way to distinguish between the torus-fed and wind-fed models.[Not sure about the text below. Maybe stop here in the Conclusion and move the rest to the Discussion section?]  

Second, our analysis demonstrates that the wind-fed models are more appropriate for Sgr~A* than those that are fed by tori. 
If wind-fed plasma circularizes at large enough distances it is possible that it would mimic torus-fed plasma in the innermost regions, where all $230\ \ghz$ emission is produced. Thus, a key property of our Sgr A* wind-fed models  (and thus by inference the Galactic Center accretion flow itself) is that the gas does \emph{not} circularize at large radii as discussed in \citet{Ressler2020a}.  
If the parcels of plasma in the inner tens of $\rg$ do indeed have the broad distribution of angular momenta (i.e., inclinations and eccentricities) that results from stochastic feeding by multiple wind sources rather than the narrow distribution provided by a torus, then certain properties of the accretion flow may never match models that consider only torus initializations.

\section*{Acknowledgements}

LM acknowledges the support of William D. Loughlin and Corning Glass Foundation memberships at the Institute for Advanced Study.
SMR was supported by the Gordon and Betty Moore Foundation through Grant GBMF7392. This research was supported in part by the National Science Foundation (NSF) under Grant No. NSF PHY-1748958.

This paper makes use of the following ALMA data:
  ADS/JAO.ALMA\#2018.1.01124.S. ALMA is a partnership of ESO (representing
  its member states), NSF (USA) and NINS (Japan), together with NRC
  (Canada) and NSC and ASIAA (Taiwan) and KASI (Republic of Korea), in
  cooperation with the Republic of Chile. The Joint ALMA Observatory is
  operated by ESO, AUI/NRAO and NAOJ.

The National Radio Astronomy Observatory is a facility of the National Science Foundation operated under cooperative agreement by Associated Universities, Inc.

This work used the Extreme Science and Engineering Discovery Environment (XSEDE) clusters Stampede2 (at the Texas Advanced Computing Center, TACC, through allocations AST170012 and AST200005) and Comet (at the San Diego Supercomputing Center, SDSC, through allocation AST090038), as well as the Princeton Research Computing cluster Stellar managed and supported by the Princeton Institute for Computational Science and Engineering (PICSciE) and the Office of Information Technology's High Performance Computing Center and Visualization Laboratory at Princeton University. Finally, this work made use of computing time granted by UCB on the Savio cluster.

\software{\texttt{grtrans} \citep{Dexter2016,Dexter2009}, \texttt{Athena++} \citep{Stone2020,White2016}}

\bibliography{sgra_variability_obs_vs_sim}{}
\bibliographystyle{aasjournal}

\appendix

\section{Structure Functions of Simplest Harmonic Sources}\label{app:model}

\begin{figure}
\begin{centering}
\includegraphics[width=0.45\columnwidth]{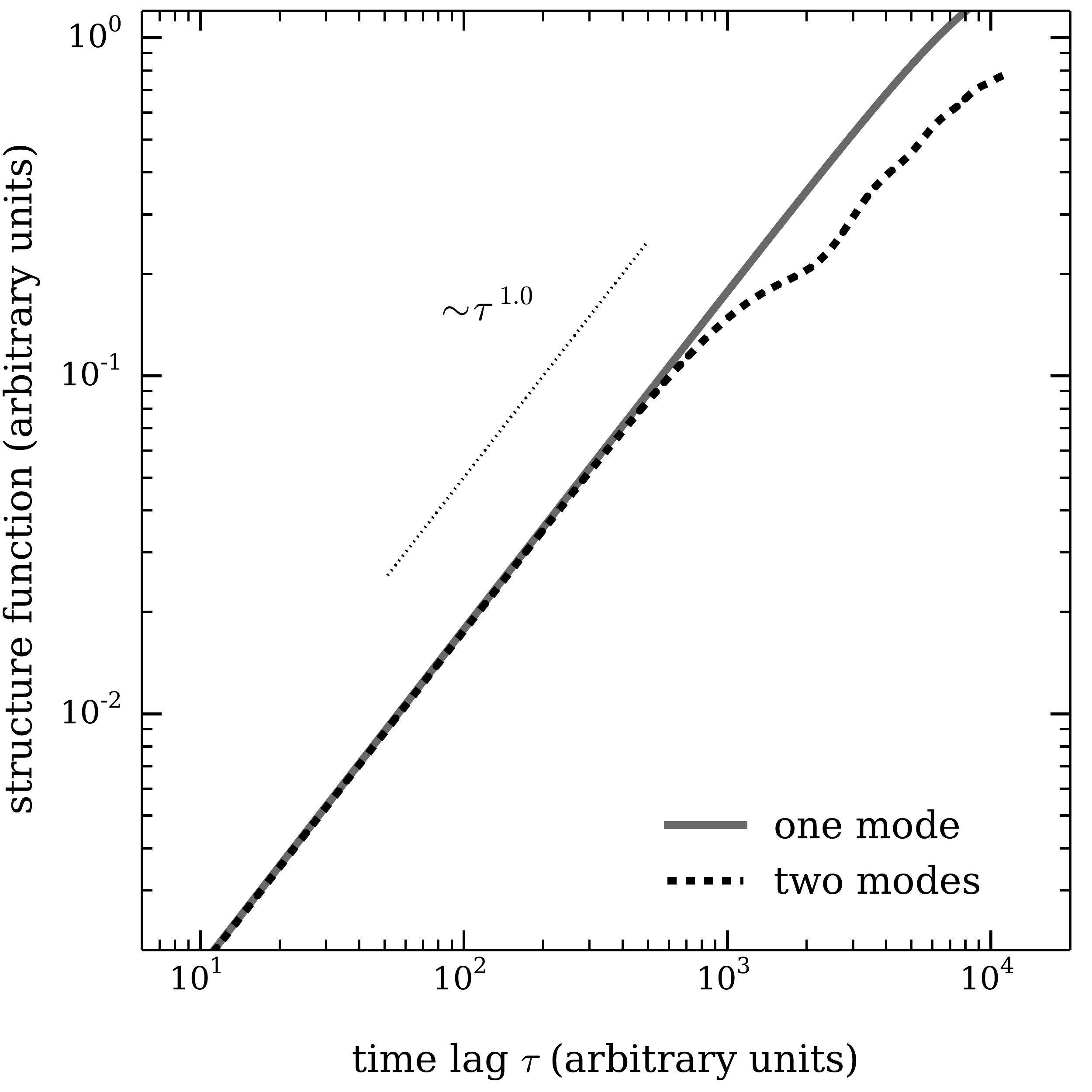}
\caption{Structure functions of simple harmonics sources described by Equation~\ref{eq:two_modes}. Solid line: Single-mode source with $A=1$, $\w = 2\pi/ 50{,}000$, $B=\nu=0$. Dashed line: Two-mode source with the same $A, \, \w$ as above and $B=0.15$, $\nu=2\pi/5000$. The $\mathcal{SF}$ is multiplied by the factor 0.55 to align with single-mode $\mathcal{SF}.$ Universal scaling ${\sim}\tau^{1.0}$ is plotted with dotted line.
}
\label{fig:model}
\end{centering}
\end{figure}

Let us calculate the structure function of a source of a single frequency  harmonic emission: $F(t)=A \sin(\w t).$ We find that 
\be
\mathcal{SF}^2(\tau)&=&\frac{A^2}{T-\tau}\int\limits^T_0 dt \left( \sin(\w (t+\tau)) -\sin(\w t) \right)^2={2 A^2}\sin^2(\w \tau),\\
\mathcal{SF}(\tau)&=&\sqrt{2} A |\sin(\w \tau)|.
\ee
Here $\w$ is the frequency of the source, $A$ the amplitude of the signal, and $T$ the duration of the light curve analyzed.
We assume that the timescale on which we study variability is much shorter than the length of the data $\tau \ll T$, and that $T \gg 2\pi/\w.$

We see that at short timescales $\tau \to 0$ the structure function has a scaling $\mathcal{SF} \sim \tau.$ This can be proven for a general flux $F(t) = \int F(\w) e^{i\w t} d\w$ with high frequency cut off $\w_0$ and any $\tau \ll 1/\w_0.$

Let us calculate structure function emission with two distinct frequencies: $F(t)=A \sin(\w t) + B \sin(\nu t +\b).$ We find
\be
\mathcal{SF}(\tau)&=&\sqrt{2} \left[A^2 \sin^2(\w \tau) + B^2 \sin^2(\nu \tau)\right]^{1/2}.\label{eq:two_modes}
\ee
Here we also assume that the two frequencies are distinct and that length of the light curve is such that $T \gg 1/(\w-\nu).$

In Figure~\ref{fig:model} we show examples of a structure function of one-~and two-harmonic sources, plotted in the same scale as the observed and simulated structure functions in Figure \ref{fig:sf_comparison}.
The period of a dominant mode is 50,000 seconds (about 14 hours) and the period of the sub-dominant mode is 5,000 seconds (83 minutes). The subdominant harmonic has an amplitude of 0.15 times that of the dominant harmonic.
We see that both structure functions trace the universal slope of $\sim \tau^{1.0}$ at small timescales. Then at about 500 seconds the two-mode structure function starts deviating from the one-mode case (which continues tracing the universal slope almost perfectly). This deviation is caused by the presence of the lower frequency sub-dominant harmonic. 
We choose the parameters in Figure \ref{fig:model} in such a way that the two-mode structure function deviates from the one-mode one at about the same timescale as the SANE torus $\mathcal{SF}$ deviates from the MAD winds $\mathcal{SF}$ (left panel in Figure \ref{fig:sf_comparison}).

We conclude that the deviation of the structure functions obtained from simulated SANE torus and MAD torus light curves from the observed ones on short timescales (Figure \ref{fig:sf_comparison}) implies the presence in the numerical models of the emission oscillation modes with shorter periods than in the observed system. Comparison with Figure~\ref{fig:model} illustrates this.

We choose the period of the dominant mode at about 14 hours to ensure there is no turn-down behavior of the one-~and two-mode structure function within the range of $\tau$ plotted. The fact that the required period is similar to the 8-hour variability period suggested by \citet{Dexter2014} was unexpected. 

\end{document}